\documentclass[reprint,aps, prl,superscriptaddress,longbibliograaphy]{revtex4-2}
\usepackage[utf8]{inputenc}
\usepackage{amsmath, amssymb}
\usepackage{bm}
\usepackage{graphicx}
\usepackage{xcolor}
\usepackage{multirow}
\usepackage{float}
\usepackage[colorlinks, linkcolor= blue, citecolor = blue, urlcolor=blue]{hyperref}

\usepackage{braket}

\def\be{\begin{equation}}
\def\ee{\end{equation}}
\def \bea{\begin{eqnarray}}
\def \eea{\end{eqnarray}}
\def \nn{\nonumber}

\begin{document}

\title{Third-order rectification in centrosymmetric metals}

\author{Sanjay Sarkar}
\email{sanjays@iitk.ac.in}
\author{Amit Agarwal}
\email{amitag@iitk.ac.in}
\affiliation{Department of Physics, Indian Institute of Technology Kanpur, Kanpur-208016, India.}

\begin{abstract} 
Rectification, the conversion of AC fields into DC currents, is crucial for optoelectronic applications such as energy harvesting and wireless communication. However, it is conventionally absent in centrosymmetric systems due to vanishing second-order optical responses. Here, we demonstrate significant rectification and photogalvanic currents in centrosymmetric metals via third-order nonlinear optical responses, driven by finite Fermi surface and disorder-induced contributions. We unveil distinct band geometric mechanisms---including Berry curvature quadrupole, Fermi surface injection, and shift effects---and classify all symmetry-allowed rectification responses. Using graphene as an example, we illustrate rectification tunability via light polarization and helicity, enabling rectification engineering in centrosymmetric materials for energy-efficient photodetection and terahertz applications.
\end{abstract}

\maketitle
\textcolor{blue}{{\it Introduction---}}  Rectification---the conversion of an AC electromagnetic field into a DC current---plays a central role in energy harvesting, wireless communication, and terahertz detection~\cite{Kronig_Nature_29,Bass_PRL_62,Isobe_2020_Sc_Adv,nakamura_NC2017_shift, deJuan_NC2017_quan, rees_SA2020_heli, ogawa_PRB2017_shift, fregoso_PRB2017_quan, wang_SA2019_ferro, yang_AM2018_light, Nechache_NP2015_band, Zhang_Nature2019_enhan, aftab_AOM2022_bulk, Ai_TJPC2020_1, onishi_PRB2024_high}. A key example is the bulk photovoltaic effect (BPVE), where light generates a DC current by exciting the material's electronic states~\cite{belinicher_SPU1980_the, baltz_PRB1981_theor, dai_CPR2023_recent}. These effects typically arise from second-order optical conductivity, $\sigma_{abc}(0;\omega,-\omega)$, which is finite only in non-centrosymmetric materials~\cite{watanabe_PRX2021_chiral,Bhalla2022}. This excludes widely available centrosymmetric materials from rectification-based applications. Higher-order optical responses, particularly third-order, offer a way to bypass this limitation. However, third-order optical nonlinearities are primarily associated with third-harmonic generation and Kerr effects~\cite{mikhailov_EL2007_non, mikhailov_PRB2015_quantum, cheng_NJP2014_third, cheng_PRB2015_third, margulis_PLA2016_freq, mikhailov_PRB2016_quan, cheng_APLP2019_intra, Jiang_NP2018_gate}, and their role in rectification remains largely unexplored~\cite{cheng_PRB2015_third, Ikeda_PRL_2023}.

In this Letter, we demonstrate that third-order optical responses provide a robust mechanism for rectification and BPVE in centrosymmetric materials \footnote{We adopt a generalized terminology for rectification (and BPVE) to refer to DC current generation in materials on application of one or more AC electric fields (optical fields). Similar phenomena have been described using alternate terms such as `two-photon or multi-photon current injection'~\cite{cheng_PRB2015_third}, `two-color coherent current injection'~\cite{PhysRevB.84.235204}, and photocurrent from  bicircular light drives~\cite{Ikeda_PRL_2023}. Our generalized use of rectification encompasses all these cases and highlights its relevance across both optical and transport regimes.}, characterized by the response tensor $\sigma_{abcd}(0; -\omega_1, -\omega_2, \omega_1+\omega_2)$. While previous studies have examined third-order nonlinearities in graphene for current injection~\cite{cheng_PRB2015_third}, we develop a general theoretical framework that applies to all centrosymmetric metals and insulators, encompassing both optical and transport regimes. By incorporating previously overlooked Fermi surface contributions and disorder effects, we develop a generalized theory of third-order rectification responses~\cite{fregoso_PRB2019_bulk, ahn_NP2022_rieman, Zhang2023_multipole, Li_Nat2024_quan, Sankar_PRX2024_exp_bcquadrupole}. This enables a systematic classification of rectification mechanisms, including Berry curvature quadrupole, Fermi surface injection, and disorder-driven contributions. 

\begin{figure}[t!] 
    \includegraphics[width=.99\linewidth]{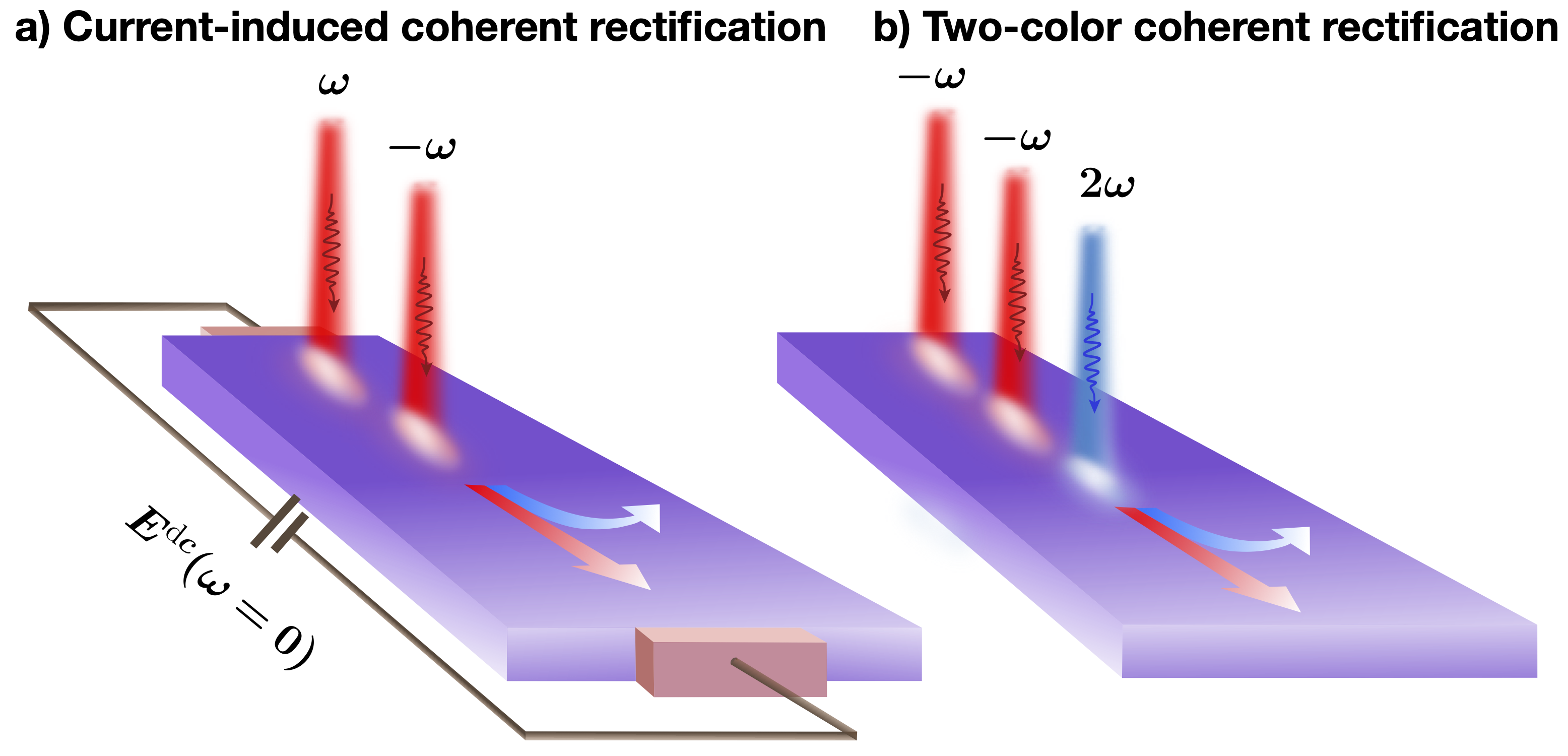}
    \caption{{\bf Schematic of third-order rectification in centrosymmetric systems}. (a) Current-induced coherent rectification (CICR) generated by a DC electric field interacting with two optical fields at frequencies $\pm \omega$. (b) Two-color coherent rectification (TCCR) driven by optical fields at frequencies $\mp \omega$ and $\pm 2 \omega$~. 
\label{Fig1}}
\end{figure}

We identify two key rectification mechanisms (see Fig.~\ref{Fig1}): (i) Current-induced coherent rectification (CICR), characterized by the response tensor $\sigma_{abcd}(0; \omega, -\omega, 0)$, and (ii) Two-color coherent rectification (TCCR), via the response tensor $\sigma_{abcd}(0; -\omega, -\omega, 2\omega)$. Both mechanisms are tunable via light polarization and helicity, generating longitudinal and transverse rectification currents. To guide material design and experimental realization, we classify all symmetry-allowed rectification responses across centrosymmetric point groups. To illustrate these concepts, we analyze polarization-tunable third-order rectification in pristine graphene, a well-characterized centrosymmetric semimetal.  Our work provides a systematic framework for engineering third-order rectification-based functionalities in centrosymmetric metals for next-generation optoelectronics, energy harvesting, and terahertz applications.

\begin{table*}[t]
	\centering
 \caption{The Fermi surface contributions to the third-order optical conductivity, $\sigma_{abcd}(-\omega_{\Sigma}; \omega_\beta, \omega_\gamma,\omega_{\delta}) = e^4/\hbar^3 \sum_{\bm k} \tilde{\sigma}_{abcd}({\bm k})$, with $\omega_{\Sigma} = \omega_\beta +  \omega_\gamma + \omega_{\delta}$. For notational simplicity, we define the complex frequencies $\tilde{\omega}_\beta=\omega_\beta+i/\tau$, $\tilde{\omega}_2=\omega_\beta + \omega_\gamma + 2i/\tau$, $\tilde{\omega}_\Sigma=\omega_\Sigma+3i/\tau$, with $\tau$ being the relaxation time. The energy bands are denoted by  $\epsilon_n ({\bm k}) = \hbar \omega_n$, and the $a$-component of the band velocity is given by $ \hbar \omega_{n;a} = v_n^a = \partial_{a} \epsilon_n ({\bm k})$ with $\partial_a \equiv \frac{\partial}{\partial k^a}$. The energy difference between two bands at a given ${\bm k}$ point is $\hbar\omega_{nm}=\epsilon_n ({\bm k})- \epsilon_m({\bm k})$, with the corresponding velocity injection being $\omega_{nm;a} = \partial_{a} \omega_{nm}$. $f_n$ denotes the Fermi function for the $n-$th band. $\Omega$ is the Berry curvature, ${\cal G}$ is the quantum metric, $\Gamma$ is the metric connection, and $\tilde{\Gamma}$ is the symplectic connection. These band geometric quantities are defined in Eqs. \eqref{QGT_eq} and \eqref{QGC_eq}. The last two columns indicate which band geometric contributions remain finite under inversion ($\mathcal P$) or time-reversal ($\mathcal T$) or combined ${\mathcal P}{\mathcal T}$ symmetry. 
 }
	{
		\begin{tabular}{c c c c c }
			\hline \hline
			\rule{0pt}{3ex}
			  Conductivity & Integrand &  $\mathcal{P}$ &  $\mathcal{T}$ or ${\mathcal P}{\mathcal T}$ \\ [1ex]
			\hline \hline
			\rule{0pt}{3ex}
			Nonlinear Drude : $\tilde{\sigma}_{abcd}^{\rm NLD}(\bm{k})$ & $\frac{i}{\tilde{\omega}_{\Sigma}\tilde{\omega}_2\tilde{\omega}_{\beta}}\sum_{n}\omega_{n;a}\frac{\partial^3 f_n}{\partial k^d \partial k^c \partial k^b}$ &  $\neq 0$ &  $\neq 0$  \\ [2ex]
			Berry curvature quadrupole : $\tilde{\sigma}_{abcd}^{\rm BCQ}(\bm{k})$ & $ \frac{1}{\tilde{\omega}_2\tilde{\omega}_{\beta}}\sum_{nm}\Omega_{mn}^{ad}\frac{\partial^2 f_n}{\partial k^c \partial k^b}$  &  $\Omega$  &  0  \\[2ex]
			Fermi surface injection-1 : $\tilde{\sigma}_{abcd}^{\rm FSI-1}(\bm{k})$ ~~& 	$-\frac{i}{\tilde{\omega}_{\Sigma}\tilde{\omega}_2}\sum_{nm}\frac{\omega_{nm;a}(\mathcal{G}_{nm}^{cb}-i\Omega_{nm}^{cb}/2)}{(\omega_{nm}-\tilde{\omega}_{\beta})}\frac{\partial f_{mn}}{\partial k^d}$ ~~& ~~ $\mathcal{G}$, $\Omega$~~ & $\mathcal{G}$ \\ [2ex]
			Fermi surface injection-2 : $\tilde{\sigma}_{abcd}^{\rm FSI-2}(\bm{k})$ &  $ \frac{i}{\tilde{\omega}_{\Sigma}} \sum_{nm}\frac{\omega_{nm;a}(\mathcal{G}_{nm}^{db}-i\Omega_{nm}^{db}/2)}{(\omega_{nm}-\tilde{\omega}_2)(\omega_{nm}-\tilde{\omega}_{\beta})} \frac{\partial f_{mn}}{\partial k^c}$  & $\cal{G}$, $\Omega$ & $\mathcal{G}$ \\[2ex]
			Fermi surface shift : $\tilde{\sigma}_{abcd}^{\rm FSS}(\bm{k})$ & $\frac{i}{\tilde{\omega}_{\beta}}\sum_{nm}\frac{(\Gamma_{mn}^{cad}-i\tilde{\Gamma}_{mn}^{cad})}{(\omega_{nm}-\tilde{\omega}_{\beta})}\frac{\partial f_{mn}}{\partial k^b}$  & $\Gamma$, $\tilde{\Gamma}$ & $\Gamma$  \\[2ex]		
						\hline \hline
	\end{tabular}}
	\label{table_1}
\end{table*}
%
\textcolor{blue}{{\it Rectification in centrosymmetric systems---}} 
In centrosymmetric systems, third-order rectification arises from the interaction of three AC (or optical) electric fields with frequencies $\{-m, -n, m+n\} \omega$. Below, we focus on rectification currents driven by two key mechanisms illustrated in Fig.~\ref{Fig1}, and discussed earlier. CICR driven by a DC electric field coupled with two optical fields at frequencies $\{0, \omega, -\omega\}$, and TCCR induced by three AC fields at frequencies $\{-\omega, -\omega, 2\omega\}$. 

To establish CICR, consider LP light incident normally on a 2D system in the $x$-$y$ plane, described by ${\bm E}(t) = e^{i\omega t} E_\omega (\cos\theta, \sin\theta, 0)$ and a DC field ${\bm E}^{\rm dc} = E_y^{\rm dc} \hat{y}$. The third-order CICR current is expressed as $j_a^{\rm LP} = \sigma_{a}^{\rm LP}|E_\omega^2|E_y^{\rm dc}$ for $a = x$ or $y$, where the effective rectification conductivity depends on the polarization angle, 
\be \label{LP}
\sigma_{a}^{\rm LP}(\theta) = {\rm Re}\left[\sigma_{axxy}\cos^2\theta + \sigma_{ayyy}\sin^2\theta + \sigma_{axyy}\sin2\theta\right].
\ee
Here, $\sigma_{abcd}$ are the third-order optical conductivity tensors, and $\theta$ captures the polarization angle of the LP light with respect to the $x$-axis. We discuss the distinct band geometric contributions to each of these conductivities in detail in the next section.  

For CP light, helicity introduces unique effects. The electric field is described as ${\bm E}(t) =e^{i\omega t} E_\omega (1, \lambda i,0)$ with $\lambda = \pm 1$ for right- or left-handed polarization. The helicity-dependent CICR current is given by $j_a^{\rm CP} = \sigma_{a}^{\rm CP} |E_\omega^2|E_y^{\rm dc}$, with the rectification conductivity,
\be \label{CP}
\sigma_a^{\rm CP} (\lambda) = {\rm Re}[\sigma_{axxy}+\sigma_{ayyy}] - 2\lambda {\rm Im}[\sigma_{axyy}].
\ee
We extend this framework to TCCR, analyzing the dependence of rectification currents on polarization angle and helicity.  
The detailed derivations for both these configurations, considering an arbitrary angle of the DC field in CICR, and the $2\omega$ field in TCCR, are presented in Sec.~S1 of the Supplementary Material (SM) \footnote{
{The Supplementary Material discusses: i) Rectification currents under different polarizations of light, ii) Light-matter interaction, iii) Calculation of density matrices, iv) Third-order optical susceptibilities, and v) Third-order Rectification responses, and vi) Current induced and two color coherent rectification}}.  
Our framework naturally extends to 3D systems, providing a robust framework for designing nonlinear optoelectronic functionalities in centrosymmetric materials. Note that tailored light-matter interactions driven photocurrents through bicircular light drives are an example of TCCR with CP light~\cite{Ikeda_PRL_2023, Trevisan_PRL_2022, Rana_PRA_2022}. 
Having established the role of polarization and helicity in third-order rectification current, we now examine the underlying optical conductivity tensors that govern these effects.





\textcolor{blue}{{\it Third-order optical response---}} 
Our approach for calculating third-order optical conductivity goes beyond previous studies by (i) treating all three electric fields coherently, (ii) retaining both Fermi surface and Fermi sea contributions, and (iii) incorporating the effect of disorder. This helps in uncovering significant rectification contributions that were previously unexplored. Fermi surface contributions are particularly important for metals and doped semimetal or semiconductors. Additionally, accounting for disorder effects, with scattering timescales $\approx 10^{-12}$s, ensures the validity of our analysis for terahertz and lower frequencies. 

The interaction of crystalline materials with an electromagnetic field in the length gauge~\cite{Sipe93, aversa_PRB1995_nonlin, Sipe99, Nastos06,virk_PRB2007_semi, Nastos10} is described by the Hamiltonian $\mathcal{H}=\mathcal{H}_{0}-\textit{e}\hat{\bm r}\cdot{\bm E}\label{eq:1}$. 
Here, $\mathcal{H}_{0}$ is the unperturbed Hamiltonian in the absence of an electromagnetic field (${\bm E}$), $e<0$ is the electronic charge, and $\hat{\bm r}$ is the position operator. The polarization ${\bm P}$ induced by ${\bm E}$ can be expanded as  \cite{Boyd20}
${\bm P}={\bm P}_0 +\chi^{(1)}|{\bm E}|+\chi^{(2)}|{\bm E}|^2+ \chi^{(3)}|{\bm E}|^3+\cdots~.$
Here, $\chi^{(1)}$, $\chi^{(2)}$, and $\chi^{(3)}$ are the linear, second-order, and third-order optical susceptibilities, respectively. 
%
The third-order optical susceptibility, $\chi^{(3)}$ is related to the third-order polarization by, 
\be
P_{a}^{(3)}=\sum_{b,c,d} \chi_{abcd}^{(3)}(-\omega_{\Sigma};\omega_{\beta},\omega_{\gamma}, \omega_{\delta})E^{\beta}_b E^{\gamma}_c E^{\delta}_d e^{-i\omega_{\Sigma}t}~.
\ee
Here, $\omega_\Sigma = \omega_{\beta} + \omega_\gamma + \omega_\delta$ is the sum of frequencies of the applied electric field. The polarization current is given by ${\bm J} =\frac{d{\bm P}}{dt}$. The susceptibility is related to the conductivity via the relation: $\chi_{abcd}^{(3)}(-\omega_{\Sigma};\omega_{\beta},\omega_{\gamma}, \omega_{\delta})= i \sigma^{(3)}_{abcd}(-\omega_{\Sigma};\omega_{\beta},\omega_{\gamma}, \omega_{\delta})/\omega_{\Sigma}$.

We compute the nonequilibrium polarization current using the quantum kinetic theory framework for the density matrix, incorporating disorder effects through the adiabatic switching approximation~\cite{culcer_PRB2017_inter, kumar_PRB2024_band, mandal_PRB2024_quan}. The third-order density matrix is systematically derived as a power series in the electric field strength, treating all optical fields coherently and retaining both intraband and interband contributions. A comprehensive derivation of the nonequilibrium density matrix, along with the generalized third-order optical conductivity and rectification mechanisms, is provided in Sec.~S3, S4, and S5 of the SM \cite{Note2}.

\begin{figure}[t!] 
    \includegraphics[width=\linewidth]{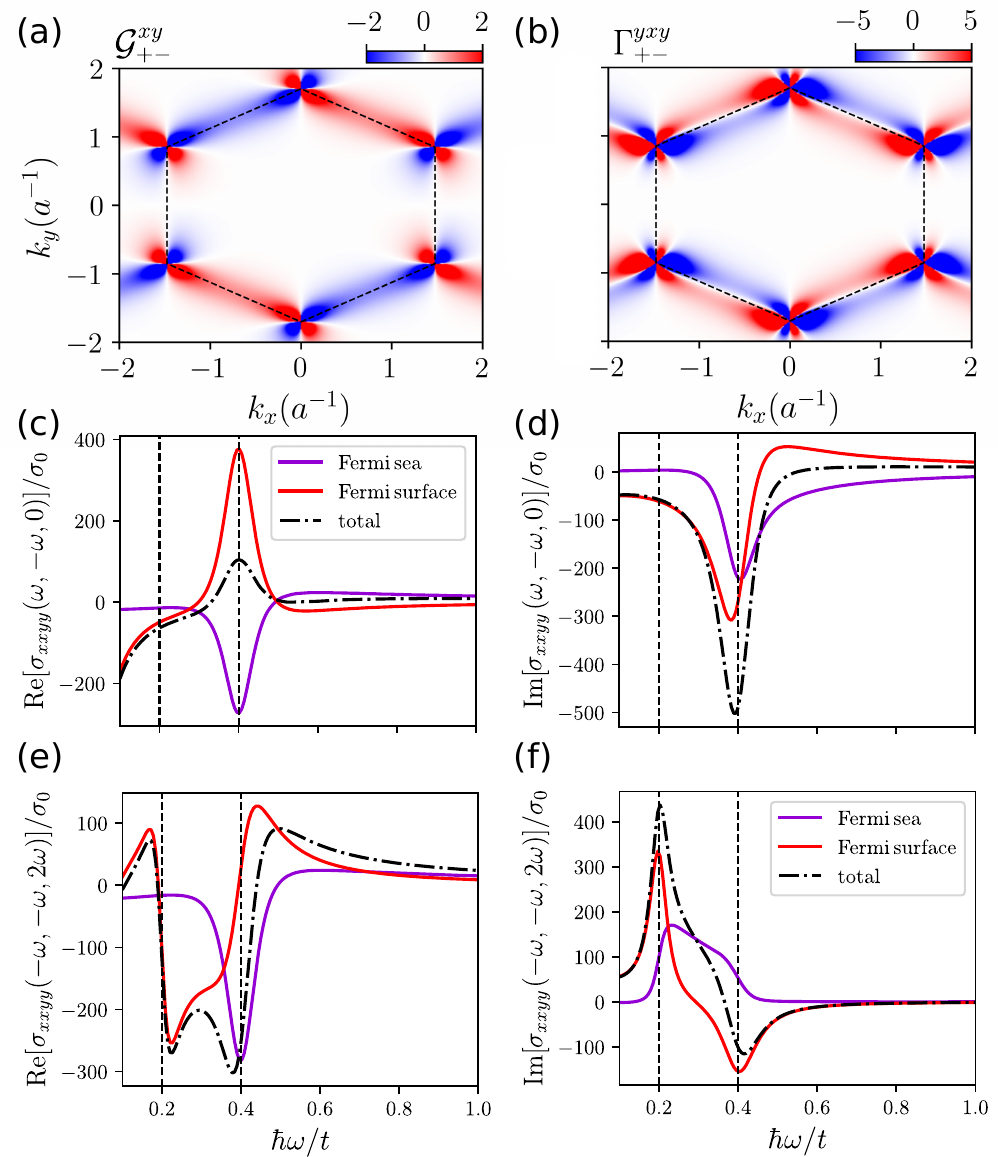}
    \caption{{\bf Fermi surface contributions dominate the third-order rectification in pristine graphene}. 
     (a) The interband quantum metric $\mathcal{G}^{xy}_{+-}$,  and (b) metric connection $\Gamma^{yxy}_{+-}$. Both of these band geometric quantities generate Fermi surface response, and the band crossing points serve as a hotspot for these. (c), (e) The real and (d), (f) the imaginary part of the two rectification responses as a function of frequency. In contrast to the Fermi sea contributions which resonances occur in vicinity of $\hbar \omega = 2 \mu$,  the Fermi surface contributions can have resonances at $\hbar \omega = \mu$ (marked by vertical line).  Conductivities are expressed in units of $\sigma_0=e^4a^2/(\hbar t^2)$. We have used $a=1.42$ ${\rm \AA}$, $t=1.0$ eV, $\mu=0.2$ eV, $\tau=1.3\times 10^{-14}$ s, and temperature $T=30$ K. 
\label{Fig2}}
\end{figure}

Our detailed calculations show that the third-order optical conductivity comprises two distinct sets of contributions: Fermi surface and Fermi sea terms, $\sigma_{abcd} = \sigma^{\rm Surface}_{abcd} + \sigma^{\rm Sea}_{abcd}$. While the Fermi sea contributions have been studied earlier \cite{fregoso_PRB2019_bulk, ahn_NP2022_rieman}, the Fermi surface contributions remain largely unexplored. However, as shown in Figs.~\ref{Fig2}(c)-(f) for graphene, these Fermi surface contributions play a pivotal role in metallic and doped semimetallic systems. Notably, they dominate the total third-order optical responses in a wide frequency range and are crucial for understanding nonlinear optical responses. 

We identify five distinct Fermi surface contributions: (i) Nonlinear Drude (NLD), (ii) BCQ, (iii) two Fermi-Surface Injection terms (FSI-1 and FSI-2), and a (iv) Fermi-Surface Shift (FSS) term. Combining these we have $\sigma^{\rm Surface}_{abcd} = \sigma^{\rm NLD}_{abcd} + \sigma^{\rm BCQ}_{abcd} + \sigma^{{\rm FSI}-1}_{abcd} + \sigma^{{\rm FSI}-2}_{abcd} + \sigma^{\rm FSS}_{abcd}$. We summarize the explicit expressions for these terms in Table~\ref{table_1}. Identifying these five Fermi surface contributions to the third-order optical conductivity is one of the significant results of our work. 

Interestingly, we find that all band geometric contributions~\cite{ahn_NP2022_rieman, lahiri_PRB2023_int, das_PRL2024_non, Adak_NRM2024_tuna, ghorai_PRL2025_planar} to the third-order optical conductivities can be expressed in terms of four key band geometric quantities: quantum metric ($\mathcal{G}_{mp}^{ab}$), Berry curvature ($\Omega_{mp}^{ab}$), metric connection ($\Gamma_{mp}^{abc}$), and the symplectic connection ($\tilde{\Gamma}_{mp}^{abc}$). These represent the real and imaginary components of two complex geometric tensors: the quantum geometric tensor ($\mathcal{Q}_{mp}^{ab}$) \cite{Bhalla2022, bhalla_PRB2023_quan} and the geometric connection ($\mathcal{C}_{mp}^{abc}$) \cite{ahn_PRX2020_low, watanabe_PRX2021_chiral}, defined as
\bea \label{QGT_eq}
\mathcal{Q}_{mp}^{ab} &=& \mathcal{R}_{pm}^a \mathcal{R}_{mp}^b = \mathcal{G}_{mp}^{ab} - \frac{i}{2} \Omega_{mp}^{ab}~,
\\
\label{QGC_eq}
\mathcal{C}_{mp}^{abc} &=& \mathcal{R}_{pm}^a \mathcal{R}_{mp;b}^c = \Gamma_{mp}^{abc} - i\tilde{\Gamma}_{mp}^{abc} ~.
\eea
Here, $\mathcal{R}_{mp}^a$ is the $a-$th component of the band resolved Berry connection, and $\mathcal{R}_{mp;b}^c=\partial_b \mathcal{R}_{mp}^c-i(\mathcal{R}_{mm}^b-\mathcal{R}_{pp}^b)\mathcal{R}_{mp}^c$ is the covariant derivative of $\mathcal{R}_{mp}^c$ with respect to Bloch momentum $k_b$. Figures~\ref{Fig2}(a) and (b) illustrate the momentum dependence of the key band geometric quantities that are significant for Fermi surface-driven third-order rectification responses in graphene. These band geometric quantities are localized near the Dirac points and are crucial for understanding the nonlinear optical responses. 

We present a comprehensive derivation of all Fermi surface contributions in Sec.~S4 and S5 of the SM~\cite{Note2}. 
Figure~S2 of the SM~\cite{Note2} illustrates the distinct Fermi surface contributions to both CICR and TCCR responses. 
As shown in Fig.~\ref{Fig2} and Fig.~S2, the resonance characteristics of Fermi sea and Fermi surface contributions exhibit clear differences. While the Fermi sea contributions feature a resonance at $\hbar\omega = 2\mu$, the Fermi surface contributions can exhibit resonances at both $\hbar\omega = \mu$ and $\hbar\omega = 2\mu$. This distinction provides an experimental signature for differentiating between Fermi sea and Fermi surface contributions. 
 

In addition to the Fermi surface contributions discussed above, the third-order optical conductivity includes other known Fermi sea contributions~\cite{Sipe90,fregoso_PRB2019_bulk}. We summarize these Fermi sea contributions systematically in Table S1 of the SM~\cite{Note2}, and identify four distinct contributions: jerk current~\cite{fregoso_PRL2018_jerk}, injection current, combined injection and shift current, and shift current. 
The third-order shift current is induced by another band geometric quantity, the second-order connection. 
In the $\omega \tau \ll 1$ limit, the third order jerk, injection, and shift currents scale with the relaxation time as $\tau^{2}$, $\tau^{1}$, and $\tau^{0}$, respectively, indicating their distinct physical origins.  
The total third-order rectification responses are obtained by combining the Fermi sea (Table S1 of the SM \cite{Note2}) and Fermi surface contributions (Table~\ref{table_1}), both of which include disorder effects. 

\textcolor{blue}{{\it Rectification in pristine graphene---}} To illustrate our results, we calculate the CICR and TCCR current in pristine monolayer graphene, a widely studied centrosymmetric system. The tight-binding Hamiltonian for graphene, considering only nearest-neighbor hopping, is a $2 \times 2$ matrix with elements $\mathcal{H}_{11} = \mathcal{H}_{22} = 0$, and $\mathcal{H}_{12} = \mathcal{H}^*_{21} = t f(\bm{k})$. Here, $t$ is the hopping parameter and $f(\bm{k})=e^{-ik_x a}\left[1+2 e^{i3k_x a/2} \cos \left(k_y a\sqrt{3}/2\right)\right]$, with $a$ being the lattice constant. Using this Hamiltonian, we calculate the band geometric quantities and present the momentum space distribution of $\mathcal{G}^{xy}_{+-}$ and $\Gamma^{yxy}_{+-}$ in Figs.~\ref{Fig2}(a) and (b). Here, $+(-)$ denotes the conduction (valence) band. Both the Berry curvature and the symplectic connection vanish due to the combined presence of inversion and time reversal symmetry in graphene. Fig.~S3 of the SM~ \cite{Note2} shows additional band geometric quantities. All the band geometric quantities are prominent near the Dirac points at the $K$ and $K'$ valleys. 

Our crystalline symmetry analysis reveals that TCCR and the $\mathcal{T}$-even CICR responses have only three independent components for graphene and other materials in the $6/mmm$ point group. These are $\sigma_{xxxx} = \sigma_{yyyy}$, $\sigma_{xxyy} = \sigma_{yxyx}=\sigma_{xyxy}=\sigma_{yyxx}$, and $\sigma_{xyyx}=\sigma_{yxxy}=\sigma_{xxxx}-2\sigma_{xxyy}$. In contrast, the $\mathcal{T}$-odd CICR response has only one independent component: $\sigma_{xxyy}=-\sigma_{yxyx}=-\sigma_{xyxy}=\sigma_{yyxx}$. 
Figures~\ref{Fig2}(c)-(f) illustrate the frequency dependence of the real and imaginary parts of the CICR conductivity $\sigma_{xxyy}(\omega, -\omega, 0)$ and the TCCR conductivity  $\sigma_{xxyy}(-\omega, -\omega, 2\omega)$. Notably, we find that previously ignored Fermi surface contributions dominate the total conductivity in all cases. These contributions are crucial for understanding nonlinear rectification in centrosymmetric metals, where they play a crucial role. Additionally, we find that prominent resonant features arise in the rectification conductivities for $\hbar \omega\approx \mu$ and $\hbar\omega \approx2\mu$, as seen in Fig.~\ref{Fig2}.  

\begin{figure}[t!] 
    \includegraphics[width=\linewidth]{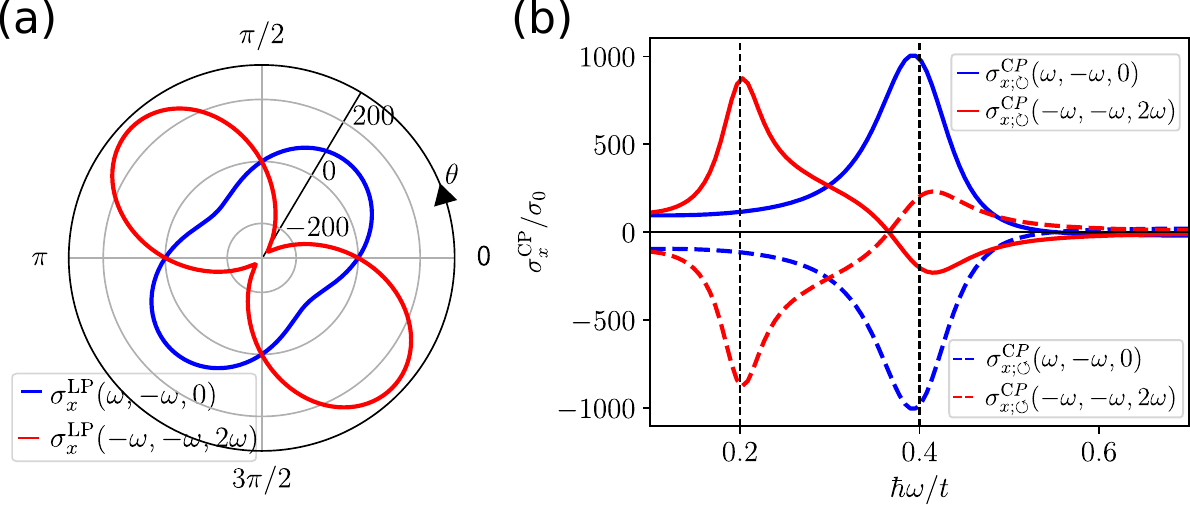}
    \caption{{\bf Polarization angle and helicity dependent third-order rectification current in graphene}. (a) Polarization angle dependence of the effective CICR and TCCR rectification conductivity for linearly polarized light at $\hbar\omega=2\mu$. The polarization angle controls both the magnitude and the direction of the rectification current. (b) The helicity and frequency dependence of the effective CICR and TCCR  conductivity for circularly polarized light. Here, $\circlearrowright$ ($\circlearrowleft$) denotes the right- (left-) handed polarization. Remarkably, the direction of both the CICR and TCCR contributions can be reversed by reversing the helicity of light. For the CICR configuration, the DC electric field is applied in the $y$ direction, and other parameters are the same as in Fig.~\ref{Fig2}. 
\label{polarization_plot}}
\end{figure}

 We now examine the polarization angle and helicity dependence of the effective rectification conductivities for LP and CP light, as defined in Eqs.~\eqref{LP}-\eqref{CP}. Figure~\ref{polarization_plot}(a) demonstrates the optical control over the magnitude and direction of the rectification current along $x$-direction for both CICR and TCCR mechanisms at $\hbar\omega=2\mu$, achieved by varying the polarization angle of the LP light. Both responses exhibit a characteristic $\sin 2\theta$ dependence, as the crystalline symmetries of graphene force $\sigma_{xxxy}$ and $\sigma_{xyyy}$ to vanish. In Fig.~\ref{polarization_plot}(b), we present the frequency dependence of the rectification current for left- and right-handed CP light. Notably, the helicity of the CP light enables a directional switch in the rectification current for both mechanisms. The corresponding helicity and frequency dependency of the rectification current along the $y$-direction are presented in Fig.~S4 of the SM~\cite{Note2}. 

From Figs.~\ref{Fig2}(c) and (e), we extract the third-order conductivity values: $\sigma_{xxyy}(\omega,-\omega,0) \simeq 0.2$ Amp.$\rm{\AA}^2/$V$^3$ and $\sigma_{xxyy}(\omega,-\omega,2\omega) \simeq 0.05$ Amp.$\rm{\AA}^2/$V$^3$ for $\hbar\omega=0.4$ eV. Using reasonable experimental parameters,   optical field strength of $10^5$ V/m, a DC field strength of similar magnitude, a sample resistance of $R \simeq 1~{\rm k}\Omega$, and a lateral sample width of $L \sim 100~\mu$m, our theoretical predictions yield rectified DC voltages of $V^{\rm{dc}} \simeq 0.2~\mu$V for CICR 
and $V^{\rm{dc}} \simeq 0.05~\mu$V for TCCR. 
These estimates align well with experimental observations of second-order rectified voltages in similar systems, which are typically of the order of $\mu$V~\cite{kumar_NN2021_room}, or smaller. This highlights the feasibility of observing third-order rectification in centrosymmetric materials and using them for energy harvesting and other optoelectronic applications.

\textcolor{blue}{{\it Crystalline symmetry restrictions---}} 
To guide experimental exploration and material design beyond graphene, we systematically analyze rectification responses in centrosymmetric materials through a comprehensive crystalline point group symmetry classification. Specifically, we examine symmetry constraints on third-order CICR and TCCR response tensors across all 11 centrosymmetric point groups. Each finite-frequency rectification response is decomposed into time-reversal even ($\mathcal{T}$-even) and time-reversal odd ($\mathcal{T}$-odd) contribution, with further details provided in the End Matter. 
We catalog all symmetry-allowed TCCR (both $T$-even and odd) and $\mathcal{T}$-even CICR response tensors for the 11 centrosymmetric point groups in Table~\ref{table_TCCR} (End Matter). Likewise, Table~\ref{table_CICR} catalogs the $\mathcal{T}$-odd CICR rectification response tensors across the same centrosymmetric point groups.
This systematic symmetry classification provides a useful framework for identifying materials with specific third-order rectification responses, enabling their application in rectification-based optoelectronic technologies.

\textcolor{blue}{{\it Conclusion---}} 
We have demonstrated that rectification currents, traditionally associated with non-centrosymmetric materials, can emerge in centrosymmetric metals via third-order optical responses. By incorporating finite Fermi surface and disorder effects, we identify distinct rectification mechanisms---including non-resonant Berry curvature quadrupole contributions, resonant Fermi surface injection, and shift contributions---that enable bulk photovoltaic effects. We highlight current-induced coherent rectification (CICR) and two-color coherent rectification (TCCR) as viable pathways for optically tunable rectification currents controlled by the polarization angle and helicity of light. Additionally, we classify all symmetry-allowed CICR and TCCR responses across centrosymmetric point groups, establishing a symmetry-based roadmap for material design and experimental realization. 

Our findings open new avenues for utilizing centrosymmetric materials in energy harvesting and terahertz photodetection through engineered third-order rectification. 
Beyond rectification, our quantum geometric framework for nonlinear responses inspires further exploration of other novel optical and transport phenomena.

\section{Acknowledgements}
We thank Koushik Ghorai for Figure 1 and fruitful discussions. We thank Kamal Das, Debottam Mandal, Sunit Das, and Sayan Sarkar, for many exciting discussions and feedback on the initial draft of the manuscript. S. S. thanks the MHRD, India, for funding through the Prime Minister's Research Fellowship (PMRF) scheme. A. A. acknowledges funding from the Core research grant by ANRF (Sanction No. CRG/2023/007003), Department of Science and Technology, India. 

\bibliography{Ref.bib}

\begin{widetext}
{\center {\bf END MATTER: Appendix}} \\

\textcolor{blue}{\it Symmetry classification of TCCR and CICR responses:---}
\begin{table*}[]
    \centering
    \begin{ruledtabular}
     \caption{Classification of all third-order TCCR responses, $\sigma_{abcd}(0; -\omega, -\omega, 2 \omega)$, and all time-reversal even (${\cal T}$-even) CICR responses, $\sigma_{abcd}(0; \omega, -\omega, 0)$, allowed by crystalline symmetries across the 11 centrosymmetric point groups. Notably, the crystalline symmetry constraints on both $\mathcal{T}$-even and $\mathcal{T}$-odd TCCR contributions are identical, and these constraints also match those for the $\mathcal{T}$-even CICR response. Additionally, we provide the relationships between non-independent response tensors.  
    \label{table_TCCR}}
    \begin{tabular}{cc}
    MPGs & {\rm TCCR} and ${\cal T}$-even CICR
    \\ 
     \hline
    $-1$ &  All components are allowed.  
    \\ 
    \hline
    $2/m$ & $xxxx$, $xyyx$, $xxyy$, $yxyx$,$yxxy$, $yyyy$, $xzzx$, $xxzx$, $xyzy$, $xyyz$, $xzzz$, $xxzz$, $yyzx$, \\ & $yzzy$, $yxzy$, $yyzz$, $yxyz$, $zxxx$, $zyyx$, $zzzx$, $zxzx$, $zyzy$, $zxyy$, $zxxz$, $zyyz$, $zxzz$, $zzzz$ 
    \\ 
    \hline
  $mmm$ & $xxxx$, $xyyx$, $xxyy$, $yxyx$,$yxxy$, $yyyy$, $xzzx$, $xxzz$, $yzzy$, $yyzz$, $zxzx$, $zyzy$, $zxxz$, $zyyz$, $zzzz$
    \\ 
    \hline
    $4/m$ & $xxxx=yyyy$, $xyyx=yxxy$, $xxyy=yxyx$, $xyyy=-yxxx$, $xxxy=-yyyx$, $xxyx=-yxyy$, $xzzx=$ \\ & $yzzy$, $xzzy=-yzzx$, $xyzz=-yxzz$, $xxzz=yyzz$, $zxzy=-zyzx$, $zyzy=zxzx$, $zxxz=zyyz$, $zzzz$
    \\ 
    \hline
    $4/mmm$ & $xxxx=yyyy$, $xyyx=yxxy$, $xxyy=yxyx$, $xzzx=yzzy$, $xxzz=yyzz$, $zyzy=zxzx$, $zxxz=zyyz$, $zzzz$
    \\ 
    \hline
    $-3$ & $xxxx=yyyy$, $xxyx=-yxyy$, $xxyy=yxyx$, $xxxy=-yyyx$, $xyyx=xxxx-2xxyy$,
    \\
    & $xyyy=xxxy+2xxyx$, $yxxx=-xxxy-2xxyx$, $yxxy=xxxx-2xxyy$, $xzzx=yzzy$, \\ & $xyzx=xxzy=yxzx=-yyzy$, $xxzx=-xyzy=-yyzx=-yxzy$, $xzzy=-yzzx$, \\ & $xxxz=-xyyz=-yxyz$, $xyzz=-yxzz$, $xxzz=yyzz$, $xxyz=yxxz=-yyyz$, $zxxx=-zyyx$, \\ & $zxxy=zxyx$, $zyzy=zxzx$, $zxzy=-zyzx$, $zxxz=zyyz$, $zzzz$
    \\ 
    \hline
    $-3m$ & $xxxx=yyyy$, $xyyx=xxxx-2xxyy$, $xxyy=yxyx$, $yxxy=xxxx-2xxyy$, $xzzx=yzzy$, \\  & $xyzx=xxzy=yxzx=-yyzy$, $xxzz=yyzz$, $xxyz=yxxz=-yyyz$ \\ & $zxxy=zxyx=-zyyy$, $zyzy=zxzx$, $zxxz=zyyz$, $zzzz$
     \\ 
    \hline
    $6/m$ & $xxxx=yyyy$, $xxyx=-yxyy$, $xxyy=yxyx$, $xxxy=-yyyx$, $xyyx=xxxx-2xxyy$,
    \\
    & $xyyy=xxxy+2xxyx$, $yxxx=-xxxy-2xxyx$, $yxxy=xxxx-2xxyy$, $xzzx=yzzy$, \\ &  $xzzy=-yzzx$, $xyzz=-yxzz$, $xxzz=yyzz$, $zyzy=zxzx$, $zxzy=-zyzx$, $zxxz=zyyz$, $zzzz$
    \\ 
    \hline
    $6/mmm$ & $xxxx=yyyy$, $xxyy=yxyx$, $xyyx=yxxy=xxxx-2xxyy$, $xzzx=yzzy$, \\ & $xxzz=yyzz$, $zyzy=zxzx$, $zxxz=zyyz$, $zzzz$
    \\ 
    \hline
    $m-3$ &  $xxxx=yyyy=zzzz$, $xxyy=yyzz=zxzx$, $xyyx=yzzy=zxxz$, $xzzx=yxxy=zyyz$, \\ & $xxzz=yxyx=zyzy$
    \\ 
    \hline
    $m-3m$ & $xxxx=yyyy=zzzz$, $xyyx=yxxy=xzzx=yzzy=zxxz=zyyz$, $xxzz=xxyy=yxyx$ \\ & $=yyzz=zxzx=zyzy$
    \\
\end{tabular}
\end{ruledtabular}
\end{table*}
%
This appendix provides a detailed symmetry classification of third-order TCCR and CICR responses in centrosymmetric materials. Under time-reversal symmetry ($\mathcal{T}$), the CICR conductivity satisfies the following symmetry relation~\cite{Gallego2019_actacrystal_symmetry_bilbao}, 
\be 
\mathcal{T}\sigma_{ijkl}(0;\omega,-\omega,0)=\sigma_{ikjl}(0;\omega,-\omega,0)
\ee 
Using this, we define time-reversal even ($\mathcal{T}$-even) and time-reversal odd ($\mathcal{T}$-odd) CICR response functions as: 
\bea 
\sigma_{ijkl}^{\rm even}(0;\omega,-\omega,0)&=&\frac{1}{2}\left[\sigma_{ijkl}(0;\omega,-\omega,0) + \sigma_{ikjl}(0;\omega,-\omega,0)\right],\nn\\~
\sigma_{ijkl}^{\rm odd}(0;\omega,-\omega,0)&=&\frac{1}{2}\left[\sigma_{ijkl}(0;\omega,-\omega,0) - \sigma_{ikjl}(0;\omega,-\omega,0)\right]. \\
\eea 
By construction, these $\mathcal{T}$-even and $\mathcal{T}$-odd CICR responses satisfy: 
\be 
\mathcal{T} \sigma_{ijkl}^{\rm even}(0;\omega,-\omega,0)  =   \sigma_{ijkl}^{\rm even}(0;\omega,-\omega,0)~,~~{\rm and}~~~ 
\mathcal{T} \sigma_{ijkl}^{\rm odd}(0;\omega,-\omega,0)  =  -\sigma_{ijkl}^{\rm odd}(0;\omega,-\omega,0)~.
\ee
The Jahn symbols for these responses are obtained following the recipe in Ref.~\cite{Gallego2019_actacrystal_symmetry_bilbao}. For the $\mathcal{T}$-even CICR response tensor, we calculate the Jahn symbol to be $V[V2]V$, while for the $\mathcal{T}$-odd CICR response tensor, it is $aV\{V2\}V$. Both CICR response tensors are symmetric under the pairwise exchange of the first two indices.

For the TCCR response, the time-reversal symmetry relation is given by, 
\be 
\mathcal{T}\sigma_{ijkl}(0;-\omega,-\omega,2\omega)=\sigma_{ikjl}(0;\omega,\omega,-2\omega)~.
\ee 
Using this, we define $\mathcal{T}$-even and $\mathcal{T}$-odd TCCR response functions as
\bea 
\sigma_{ijkl}^{\rm even}(0;-\omega,-\omega,2\omega)=\frac{1}{2}\left[\sigma_{ijkl}(0;-\omega,-\omega,2\omega) + \sigma_{ikjl}(0;\omega,\omega,-2\omega)\right]~,
\eea 
\bea 
\sigma_{ijkl}^{\rm odd}(0;-\omega,-\omega,2\omega)=\frac{1}{2}\left[\sigma_{ijkl}(0;-\omega,-\omega,2\omega) -\sigma_{ikjl}(0;\omega,\omega,-2\omega)\right]~.
\eea 
These responses satisfy the following relations, 
\be 
\mathcal{T} \sigma_{ijkl}^{\rm even}(0;-\omega,-\omega,2\omega)= \sigma_{ijkl}^{\rm even}(0;-\omega,-\omega,2\omega)~,~~{\rm and}~~\mathcal{T} \sigma_{ijkl}^{\rm odd}(0;-\omega,-\omega,2\omega)= -\sigma_{ijkl}^{\rm odd}(0;-\omega,-\omega,2\omega)~.
\ee 
For the TCCR responses, the Jahn symbol for the $\mathcal{T}$-even part is $V[V2]V$, while for the $\mathcal{T}$-odd part, it is $aV[V2]V$.

Using these Jahn symbols, we classify all symmetry-allowed responses for the 11 centrosymmetric point groups. The results for TCCR and $\mathcal{T}$-even CICR responses are summarized in Table~\ref{table_TCCR}, while those for $\mathcal{T}$-odd CICR responses are presented in Table~\ref{table_CICR}. This symmetry classification serves as a practical tool for identifying candidate materials and guiding experimental efforts to uncover third-order rectification phenomena.

\begin{table*}[]
    \centering
    \begin{ruledtabular}
     \caption{Classification of all third-order time-reversal odd (${\cal T}$-odd) CICR responses, $\sigma_{abcd}(0; \omega, -\omega, 0)$, allowed by crystalline symmetries for the 11 centrosymmetric point groups. We also provide the relationships between non-independent response tensors. 
    \label{table_CICR}}
    \begin{tabular}{cc}
    MPGs & ${\cal T}$-odd {\rm CICR}
    \\ 
     \hline
    $-1$ &  $xxyx=-xyxx$, $xxzx=-xzxx$, $xxyy=-xyxy$, $xxzy=-xzxy$, $xxyz=-xyxz$, $xxzz=-xzxz$, \\ &  $yxyx=-yyxx$, $yxzx=-yzxx$, $yxyy=-yyxy$, $yxzy=-yzxy$, $yxyz=-yyxz$, $yxzz=-yzxz$, \\ &  $zxyx=-zyxx$, $zxzx=-zzxx$, $zxyy=-zyxy$, $zxzy=-zzxy$, $zxyz=-zyxz$, $zxzz=-zzxz$, \\ &  $xyzx=-xzyx$, $xyzy=-xzyy$, $xyzz=-xzyz$, $yyzx=-yzyx$, $yyzy=-yzyy$, $yyzz=-yzyz$, \\ & $zyzx=-zzyx$, $zyzy=-zzyy$, $zyzz=-zzyz$  
    \\ 
    \hline
    $2/m$ & $xxzx=-xzxx$, $xxyy=-xyxy$, $xxzz=-xzxz$, $yxyx=-yyxx$, $yxzy=-yzxy$, $yxyz=-yyxz$, $zxzx=$ \\ & $-zzxx$, $zxyy=-zyxy$, $zxzz=-zzxz$, $xyzy=-xzyy$, $yyzx=-yzyx$, $yyzz=-yzyz$, $zyzy=-zzyy$ 
    \\ 
    \hline
  $mmm$ & $xxyy=-xyxy$, $xxzz=-xzxz$, $yxyx=-yyxx$, $zxzx=-zzxx$, $yyzz=-yzyz$, $zyzy=-zzyy$
    \\ 
    \hline
    $4/m$ & $xxyx=yxyy=-xyxx=-yyxy$, $xxyy=-yxyx=-xyxy=yyxx$, $xxzz=yyzz=-xzxz=-yzyz$, $zxzx=$ \\ & $zyzy=-zzxx=-zzyy$, $zxzy=-zyzx=zzyx=-zzxy$, $zxyz=-zyxz$, $xyzz=-yxzz=yzxz=-xzyz$
    \\ 
    \hline
    $4/mmm$ & $xxyy=-yxyx=-xyxy=yyxx$, $xxzz=yyzz=-xzxz=-yzyz$, $zxzx=zyzy=-zzxx=-zzyy$
    \\ 
    \hline
    $-3$ & $xxyx=yxyy=-xyxx=-yyxy$, $xxzx=-yxzy=-xyzy=-yyzx=-xzxx=xzyy=yzyx=yzxy$,
    \\
    & $xxyy=-yxyx=-xyxy=yyxx$, $xxzy=yxzx=xzzx=-yyzy=-xzyx=-xzxy=-yzxx=yzyy$, \\ & $xxzz=yyzz=-xzxz=-yzyz$, $zxz=zyzy=-zzxx=-zzyy$, $zxzy=-zyzx=zzyx=-zzxy$, \\ &  $zxyz=-zyxz$, $xyzz=-yxzz=-xzyz=yzxz$
    \\ 
    \hline
    $-3m$ & $xxyy=-yxyx=-xyxy=yyxx$, $xxzy=yxzx=xyzx=-yyzy=-xzyx=-xzxy=-yzxx=yzyy$, \\  & $xxzz=yyzz=-xzxz=-yzyz$, $zxzx=zyzy=-zzxx=-zzyy$
     \\ 
    \hline
    $6/m$ & $xxyx=yxyy=-xyxx=-yyxy$, $xxyy=-yxyx=-xyxy=yyxx$, $xxzz=yyzz=-xzxz=-yzyz$, $zxzx=$
    \\
    & $zyzy=-zzxx=-zzyy$, $zxzy=-zyzx=zzyx=-zzxy$, $zxyz=-zyxz$, $xyzz=-yxzz=-xzyz=yzxz$ 
    \\ 
    \hline
    $6/mmm$ & $xxyy=-yxyx=-xyxy=yyxx$, $xxzz=yyzz=-xzxz=-yzyz$, $zxzx=zyzy=-zzxx=-zzyy$ 
    \\ 
    \hline
    $m-3$ &  $xxyy=-zxzx=-xyxy=yyzz=-yzyz=zzxx$, $xxzz=-yxyx=yyxx=-zyzy=-xzxz=zzyy$ 
    \\ 
    \hline
    $m-3m$ & $xxyy=xxzz=-yxyx=-zxzx=-xyxy=yyxx=yyzz=-zyzy=-xzxz=-yzyz=zzxx=zzyy$
    \\
\end{tabular}
\end{ruledtabular}
\end{table*}
\end{widetext}


\end{document}